\documentclass[aps,prl,floatfix,twocolumn,superscriptaddress,showpacs,10pt]{revtex4-1}
\usepackage{amssymb, amsmath,color,mciteplus,graphicx,subfigure}
\usepackage{calc,epsfig,epstopdf,color,mciteplus,bm,mathrsfs}
\usepackage{times}
\usepackage{float}
\usepackage{lipsum}
\begin{document}

\title{Robust and Fast Holonomic Quantum Gates with Encoding on Superconducting Circuits}

\author{Tao Chen}
\affiliation{Guangdong Provincial Key Laboratory of Quantum Engineering and Quantum Materials, 
and School of Physics\\ and Telecommunication Engineering, South China Normal University, Guangzhou 510006, China}

\author{Pu Shen}
\affiliation{Guangdong Provincial Key Laboratory of Quantum Engineering and Quantum Materials, 
and School of Physics\\ and Telecommunication Engineering, South China Normal University, Guangzhou 510006, China}

\author{Zheng-Yuan Xue}\email{zyxue83@163.com}
\affiliation{Guangdong Provincial Key Laboratory of Quantum Engineering and Quantum Materials, 
and School of Physics\\ and Telecommunication Engineering, South China Normal University, Guangzhou 510006, China}
\affiliation{Frontier Research Institute for Physics, South China Normal University, Guangzhou 510006, China}

\date{\today}

\begin{abstract}
  High-fidelity and robust quantum manipulation is the key for scalable quantum computation. Therefore, due to the intrinsic operational robustness, quantum manipulation induced by geometric phases is one of the promising candidates. However, the longer gate time for geometric operations and more physical-implementation difficulties hinder its practical and wide applications. Here, we propose a simplified implementation of universal holonomic quantum gates on superconducting circuits with experimentally demonstrated techniques, which can remove the two main challenges by introducing the time-optimal control into the construction of quantum gates. Remarkably, our scheme is also based on a decoherence-free subspace encoding, with minimal physical qubit resource, which can further immune to error caused by qubit-frequency drift, which is regarded as the main error source for large scale superconducting circuits. Meanwhile, we deliberately design the quantum evolution to eliminate gate error caused by unwanted leakage sources. Therefore, our scheme is more robust than the conventional ones, and thus provides a promising alternative strategy  for scalable fault-tolerant quantum computation.
\end{abstract}

\maketitle


Quantum computation is believed to be a promising solution for certain hard problems \cite{qc}, which will benefit for many practical applications nowadays. Thus, the physical implementation of quantum computation has attract much attentions, especially for the superconducting quantum circuit system \cite{sqc1,sqc2,sqc3,sqc4,sqc5}, due to its fine fabrication and characterization technologies. However, the scalability of quantum computation \cite{QS2019} is challenging due to the inevitable noises and operational errors. Thus, due to the built-in noise-resilience features, quantum gates induced by geometric phases \cite{Abelian,non-Abelian,AA} have been proposed as a promising strategy to realize high-fidelity and robust quantum gates, formerly based on adiabatic cyclical evolution \cite{zanardi,AGQC1,Duan}. On the other hands, as the coherent times of quantum systems are limited, fast quantum gates are more preferable, as the decoherence effect will then induce less gate error. Thus, quantum computation based on both nonadiabatic Abelian \cite{wxb,ZSL1,UGQCZhu,NGQC,tchen} and non-Abelian geometric phases \cite{NJP,TongDM,Liu18} has been proposed. Remarkably, experimental demonstrations for elementary geometric quantum gates have been achieved on various systems \cite{exp1, exp2, DuJ, Abdumalikov35, Feng39, Zu41, AC2014,nv2017,nv20172,li2017, xuy37,yan2019, zhu2019, Xu2019}. However, the time of arbitrary geometric quantum gates are still much longer than that of the dynamical ones, and thus  lead to more decoherence-induced gate error \cite{BNHQC,Chentoc}.

Meanwhile, the challenge of implementing quantum computation with nonadiabatic non-Abelian geometric phases, i.e., nonadiabatic holonomic quantum computation (NHQC), on superconducting circuits lies in several aspects. First,  it  needs complex interaction among multiple-level systems, which is usually experimentally difficulty. Second, interaction and/or operation induced qubits' frequency drift effect is one of the main error sources for  multi-qubit lattices. Finally, gate error caused by unwanted quantum information leakage out of the qubit subspaces is also an important concern, especially for the two-qubit gate case.

Here, we propose to a robust and fast implementation of  NHQC with simplified setup and experimental accessible techniques with superconducting transmon qubits \cite{transmon1,transmon2}. Our scheme is based on a two-dimensional (2D) square lattice scenario, with experimentally demonstrated parametrically tunable coupling among the adjacent qubits \cite{Rigetti18,xli,ExpYuY}. Meanwhile, our scheme incorporates a minimal resource decoherence-free subspace (DFS) encoding \cite{dfs1,dfs2,dfs3}, which significantly simplifies Ref. \cite{TongDM} and can immune to the qubits' frequency drift error, and thus can combine the operational robust feather of geometric phase and decoherence resilience of the encoding. In addition, we deliberately design the evolution for the target quantum gates within the logical qubit subspace, which can totally eliminate the unwanted leakage errors. Furthermore, we also introduce the time-optimal control (TOC) technique \cite{TO1, TO2} into our gate construction, extending Ref. \cite{BNHQC} to the time-dependent case, where the shortest evolution path for a particularly gate can be found, thus minimize the decoherence induced gate infidelity. Finally, we numerically show that our scheme can be more robust than the  conventional NQHC, in terms of the main error sources for the superconducting qubits. Therefore, our proposal provides a promising way towards scalable fault-tolerant quantum computation.


\begin{figure}[tbp]
  \centering
  \includegraphics[width=\linewidth]{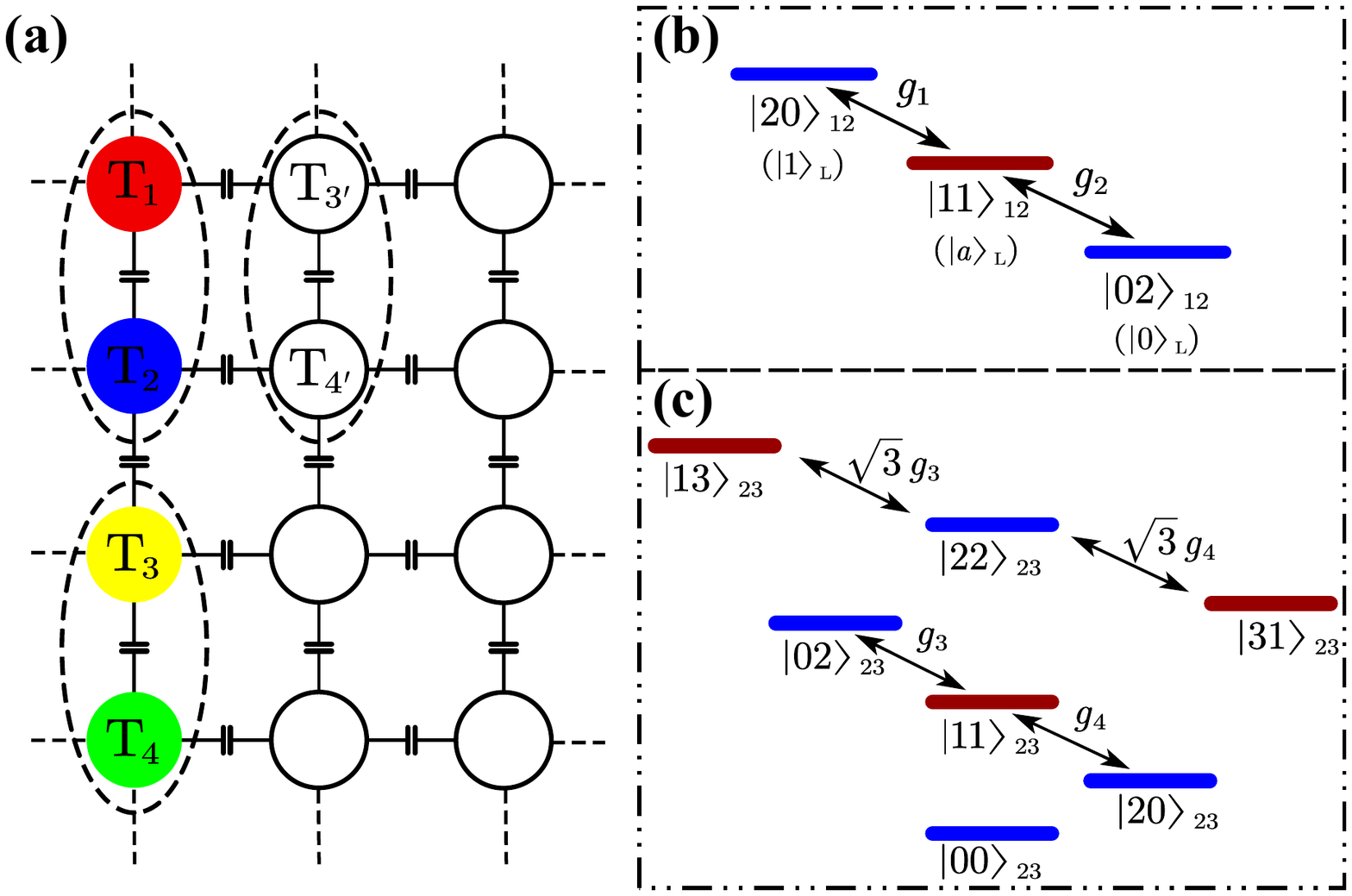}
  \caption{Illustration of our scheme. (a) A 2D square qubit lattice, with two capacitive coupled transmons being used as a logical  DFS qubit. The energy spectrum structure of two capacitive coupled transmons in the cases of (b) single- and (c) two-logical qubit.}\label{Fig1}
\end{figure}

We now present our scheme on a 2D square superconducting transmon lattice. With the requirement of the minimal qubit resource, as shown in Fig. \ref{Fig1}(a), we only use two capacitive coupled transmons $\textrm{T}_1$ and $\textrm{T}_2$ as a logical unit to encode a DFS qubit, i.e., $S_1=\{|02\rangle_{12}=|0\rangle_L, |20\rangle_{12}=|1\rangle_L\}$. Assuming that $\hbar=1$ hereafter, the corresponding Hamiltonian of the two coupled transmons reads
\begin{eqnarray}
\label{EqH12}
\mathcal{H}_{12}=&&\sum^2_{j=1}\sum^{+\infty}_{n=1}[n\omega_j-\sum^n_{k=1}(k-1)\alpha_j ]|n\rangle_{j}\langle n|  \notag \\
&&+g_{_{12}}(S_1S_2^\dagger+S^\dagger_1S_2),
\end{eqnarray}
where $S_j=\sum^{+\infty}_{n=1}\sqrt{n}|n-1\rangle_{j}\langle n|$ denotes the standard lower operator for transmon $\textrm{T}_j$; $g_{_{12}}$ is the coupling strength between two adjacent transmons $\textrm{T}_1$ and $\textrm{T}_2$; $\omega_j$ is the associated transition frequency with $\alpha_j$ being the intrinsic anharmonicity of transmon $\textrm{T}_j$. For the time-dependent tuning of the coupling, we introduce a two-tone frequency driving \cite{ExpYuY} for transmon $\textrm{T}_1$ as $\omega_1(t)=\omega_1+\varepsilon_1(t)+\varepsilon_2(t)$ with $\varepsilon_i(t)=\dot{\mathcal{F}}_i(t)$, where $\mathcal{F}_i(t)=\beta_i(t)\sin(\omega_{\varepsilon_i}t +\phi_{\varepsilon_i}(t))$ with $\omega_{\varepsilon_i}$ and $\phi_{\varepsilon_i}(t)$ being the driving frequency and phase, respectively. Then, in the interaction picture, the Hamiltonian under driven is
\begin{eqnarray}
\label{EqH12E}
\mathcal{H}_{12}^{\varepsilon}(t)
=&&\left\{ |11\rangle_{12}\langle 02|e^{\textrm{i}(\Delta_1+\alpha_2)t}
   +|20\rangle_{12}\langle 11|e^{\textrm{i}(\Delta_1-\alpha_1) t}\right\} \notag\\
&& \times \sqrt{2}g_{_{12}} \prod_{i=1,2}\left[e^{\textrm{i}\beta_i(t)\sin(\omega_{\varepsilon_i}t + \phi_{\varepsilon_i}(t))}\right]
+\mathrm{H.c.},
\end{eqnarray}
where $\Delta_1=\omega_1-\omega_2$ is the qubit-frequency difference. Note that, for the effects of qubit-frequency drifts of transmons $\textrm{T}_1$ and $\textrm{T}_2$, in the form of $\omega_1+\delta_1$ and $\omega_2+\delta_2$, the above encoding in the DFS $S_1$ can serviceably eliminate the overlapped part of qubit-frequency drifts of transmons $\textrm{T}_1$ and $\textrm{T}_2$, where the variation of the drift is of the low-frequency nature, so that $\delta_{1,2}$ can be regarded as constants during a gate. In addition, the retained drift difference $\delta_d=\delta_1-\delta_2$ can also be suppressed due to the geometric robustness, as demonstrated below.

Due to the absence of transition interactions, except for the state $|11\rangle_{12}$, other non-logical-qubit states are not involved in Eq. (\ref{EqH12E}). Therefore, by using the Jacobi-Anger identity expansion, and then modulating qubit-driving frequencies to meet $\Delta_1+\alpha_2-(\omega_{\varepsilon_1}+\omega_{\varepsilon_2}) =-(\Delta_1-\alpha_1-\omega_{\varepsilon_1})=\Delta$, with $\Delta\ll\{\omega_{\varepsilon_1}, \omega_{\varepsilon_2}\}$, under the rotating-wave approximation, see Ref.  \cite{sm} for details, Eq. (\ref{EqH12E}) forms a tunable three-level structure in the DFS with an auxiliary state of $\{|a\rangle_{L}=|11\rangle_{12}\}$, as shown in Fig. \ref{Fig1}(b), where the effective coupling strengths are $g_1(t)$ and $g_2(t)$. The distinct merit here is that, different from the non-encoding case of manipulating a single transmon qubit \cite{ME}, the unwanted leakage errors are naturally eliminated in our construction without  correction.

In the dressed-state representation $\{|\psi_+\rangle_{L},|\psi_-\rangle_{L}\}$, after a unitary transformation \cite{sm}, the dynamic process of the quantum system can also be denoted by the coupling between $|\psi_+\rangle_{L}=\cos\frac{\theta}{2}e^{\text{i}\phi}|0\rangle_{L}+\sin\frac{\theta}{2}|1\rangle_{L}$ and $|a\rangle_{L}$ with the strength $g(t)=\sqrt{g_1^2(t)+g_2^2(t)}$ and detuning $\Delta$, i.e.,
\begin{eqnarray}
\label{EqHL1}
\mathcal{H}_{L_1}(t)=-\frac{\Delta}{2}\tilde{\sigma}^z_{L}+g(t)\left(e^{-\textrm{i}\phi_2(t)}|\psi_+\rangle_{L}\langle a|+\mathrm{H.c.}\right),
\end{eqnarray}
where $\tilde{\sigma}^z_{L}=|\psi_+\rangle_{L}\langle \psi_+|-|a\rangle_{L}\langle a|$,  $\theta=2\tan^{-1}[g_2(t)/g_1(t)]$ and $\phi=\phi_2(t)-\phi_1(t)$; while state $|\psi_-\rangle_L=\sin\frac{\theta}{2}e^{\text{i}\phi}|0\rangle_L-\cos\frac{\theta}{2}|1\rangle_L$ is decoupled.  
Furthermore, defining $\xi(t)=\int\sqrt{g^2(t)+(\Delta+\dot{\phi}_2(t))^2/4} dt$, under driven Hamiltonian $\mathcal{H}_{L_1}(t)$, these two dressed states evolve as
\begin{eqnarray}
\label{EqstateE}
|\Psi_1(t)\rangle&=&U_{L_1}(t)|\psi_+\rangle_{L}=-\textrm{i} \sin\xi(t)\sin\chi e^{\textrm{i}\frac{1}{2}\phi_2(t)} |a\rangle_{L} \notag\\
&&+\left[\cos\xi(t)+\textrm{i}\sin\xi(t)\cos\chi\right]e^{-\textrm{i}\frac{1}{2}\phi_2(t)}|\psi_+\rangle_{L}, \notag\\
|\Psi_2(t)\rangle&=&U_{L_1}(t)|\psi_-\rangle_{L}=|\psi_-\rangle_{L},
\end{eqnarray}
with $\chi=\tan^{-1}[2g(t)/(\dot{\phi}_2(t)+\Delta)]$. Ensuring that $\xi(\tau_1)=\pi$ is met at a final gate-time $\tau_1$, dressed states $|\psi_+\rangle_{L}$ and $|\psi_-\rangle_{L}$ undergo cyclic evolutions, with the accumulated total phase being $\gamma_1=\pi-\phi_2(\tau_1)/2$ in state $|\psi_+\rangle_{L}$. The corresponding time-evolution operator is
\begin{eqnarray}
\label{UL1}
U_{L_1}(\tau_1)=\sum_{l,m=1,2}(\textbf{T}e^{\textrm{i}\int^{\tau_1}_0 [\textbf{A}(t)+\textbf{K}(t)] dt})_{lm}|\Psi_l(0)\rangle\langle \Psi_m(0)|, \notag
\end{eqnarray}
where $\textbf{T}$ is time-ordering operator, $A_{lm}=\textrm{i}\langle \Psi_l(t)|\frac{\partial}{\partial t}|\Psi_m(t)\rangle$ and $K_{lm}=-\langle \Psi_l(t)|\mathcal{H}_{L_1}(t)|\Psi_m(t)\rangle$ represent the geometric and dynamical elements, respectively. Here we find that generally $\textbf{K}=r\textbf{A}+\textbf{G}$ with $r\equiv-1$  \cite{BNHQC}, and matrix $\textbf{G}$ depends only on the global geometric feature of evolution path, thus
\begin{eqnarray}
\label{EqUHP}
U_{L_1}(\tau_1)=e^{\textrm{i}\textbf{G}}= \text{diag}(e^{\textrm{i}\gamma_1}, 0)
\end{eqnarray}
is an unconventional holonomy, extending the Abelian case \cite{UGQCZhu,DuJ}, which is also totally different from conventional holonomic quantum gates \cite{NJP} by taking $K_{lm}=0$ to eliminate the dynamical phase.

\begin{figure}[tbp]
  \centering
  \includegraphics[width=\linewidth]{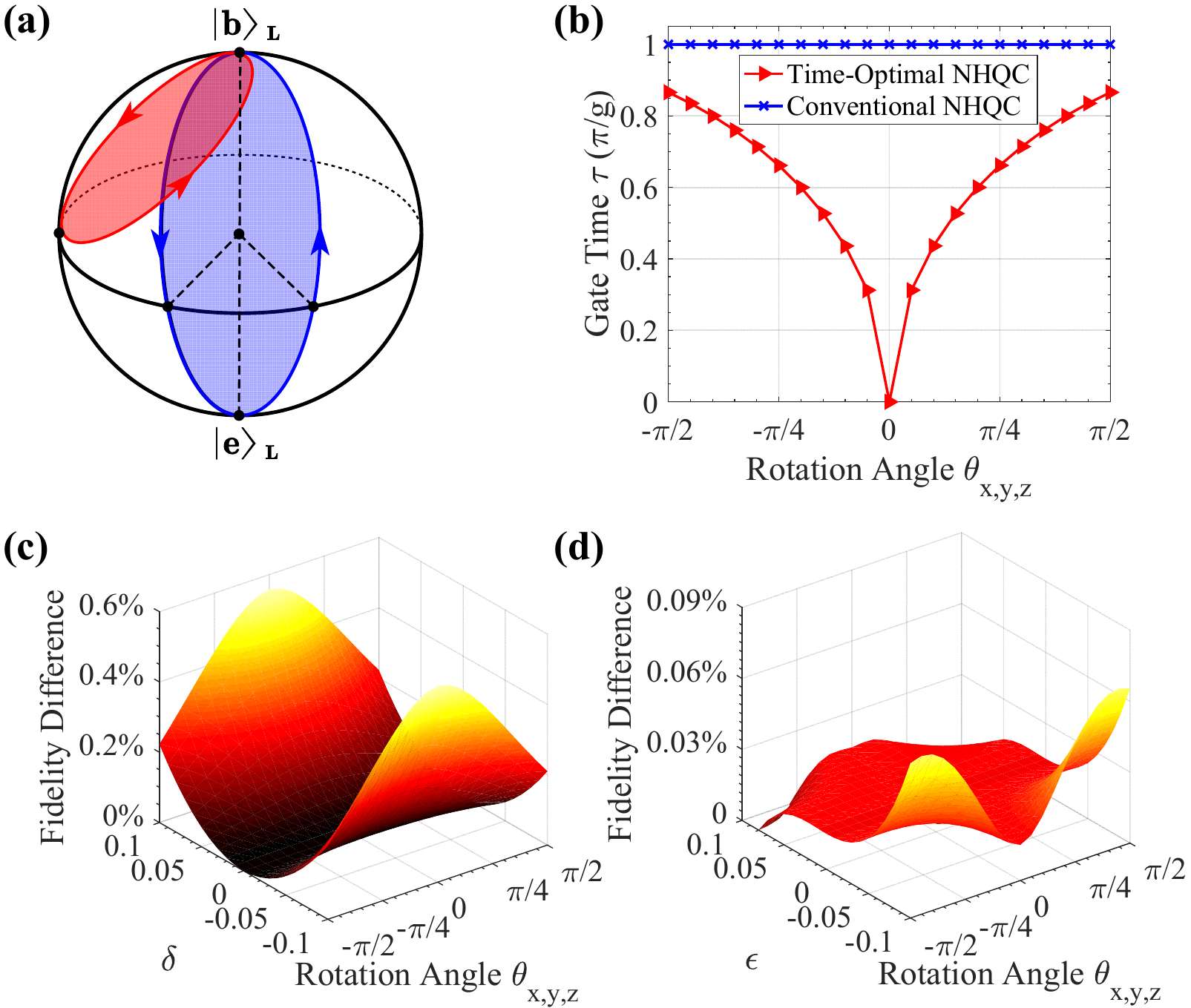}
  \caption{The construction of holonomic gates and their performance. (a) Geometric illustration of the evolution paths of our proposed time-optimal (red line) and conventional (blue line) holonomic gates in a same Bloch sphere, with results (b) of gate-time comparison at arbitrary X,Y,Z-axis rotation angles. Fidelity difference between our proposed time-optimal and conventional holonomic gates as a function of (c) frequency-drift difference $\delta_d=\delta \times g$ and (d) the deviation $\epsilon \times g$ of the coupling strength, respectively.}\label{Fig2}
\end{figure}

Based on the unconventional holonomic operation framework, tunable parameters $g(t)$ and $\phi_2(t)$, determine the evolution path of the operation, which can be shaped to accelerate holonomic gates with TOC technique. Follow Refs. \cite{TO1, TO2}, by analyzing restricted conditions of the interaction Hamiltonian $\mathcal{H}_{\textrm{int}}(t)=g(t)[\cos\phi_2(t){\tilde{\sigma}}^x_L+ \sin\phi_2(t){\tilde{\sigma}}^y_L]$ of Eq. (\ref{EqHL1}) in the realistic physical implementation, i.e., the tunable coupling strength $g(t)$ can only be adjusted within a certain range and thus not be infinite; and the form of the interaction Hamiltonian $\mathcal{H}_{\textrm{int}}(t)$ is not arbitrary, thus which can be represented, respectively, as $f_1[\mathcal{H}_{\textrm{int}}(t)]=\frac{1}{2}[\textrm{Tr}(\mathcal{H}^2_{\textrm{int}}(t))-2g^2(t)]=0$ and $f_2[\mathcal{H}_{\textrm{int}}(t)]=\textrm{Tr}(\mathcal{H}_{\textrm{int}}(t) \tilde{\sigma}^z_L)=0$, where $\tilde{\sigma}^{x,y,z}_L$ are Pauli operators in the dressed-state subspace $\{|\psi_+\rangle_L, |a\rangle_L\}$. Then, by solving the quantum brachistochrone equation \cite{QBE} $\partial\mathcal{F}/ \partial t=-i[\mathcal{H}_{L_1}(t),\mathcal{F}]$ with $\mathcal{F}=\partial (\sum_{j=1,2}\lambda_jf_j[\mathcal{H}_{\textrm{int}}(t)])/\partial \mathcal{H}_{L_1}(t)$, with $\lambda_j$ being the Lagrange multiplier, we can obtain $\phi_2(t)=\int_0^t[C_0g(t')-\Delta]dt'$, where the coefficient $C_0$ is a constant that depends only on the type of target gate.

Meanwhile, it is worth reemphasizing that since there is no need to engineer the especial shape of $g(t)$ to suppress unwanted leakage errors, we can determine $g(t)=g$ as a square pulse with corresponding $\dot{\phi}_2(t)=C_0g-\Delta=\eta_1$ being a constant to realize the shortest geometric path, as shown in Fig. \ref{Fig2}(a).  Therefore, within the single-logical-qubit subspace $S_1$, arbitrary time-optimal holonomic gates can be obtained as
\begin{eqnarray}
\label{EqU1}
U_{L_1}(\gamma_1, \theta, \phi)=\cos\frac{\gamma_1}{2}+\textrm{i}\sin \frac{\gamma_1}{2}\left(
\begin{array}{cccc}
 \cos\theta & \sin\theta e^{\textrm{i}\phi} \\
 \sin\theta e^{-\textrm{i}\phi} & -\cos\theta
\end{array}
\right). \notag\\
\end{eqnarray}
Then, by setting gate parameters $(\gamma_1, \theta, \phi)=(\theta_x, \pi/2, \pi)$, $(\theta_y, \pi/2, \pi/2)$, $(\theta_z, \pi, \pi)$, holonomic X,Y,Z-axis rotation operations $R^\textrm{T}_x(\theta_x)$, $R^\textrm{T}_y(\theta_y)$ and $R^\textrm{T}_z(\theta_z)$ can all be obtained at the optimal time $\tau_1=\tau_0\sqrt{1-(1+\Delta/\eta_1)^2(1-\gamma_1/\pi)^2}$, which are faster than conventional holonomic operations based on a single-loop scenario \cite{SingleloopSQ,Singleloop} with the gate time being $\tau_0=\pi/g$ for all gates. The corresponding acceleration effect is shown in Fig. \ref{Fig2}(b), where we set $\Delta=0$ for illustration purpose. Furthermore, we test the robustness of our holonomic operations by using $\textrm{F}^{\delta, \epsilon}=\textrm{Tr}(U^\dagger_{L_1}U^{\delta, \epsilon}_{L_1})/\textrm{Tr}(U^\dagger_{L_1}U_{L_1})$ as the fidelity formula, where $U^{\delta, \epsilon}_{L_1}$ represents the gate affected by the frequency-drift difference
and deviation  of the coupling strength. 
As shown in Figs. \ref{Fig2}(c) and \ref{Fig2}(d), our scheme shows better noise-resilient feature for both representative errors.

\begin{figure}[tbp]
  \centering
  \includegraphics[width=0.8\linewidth]{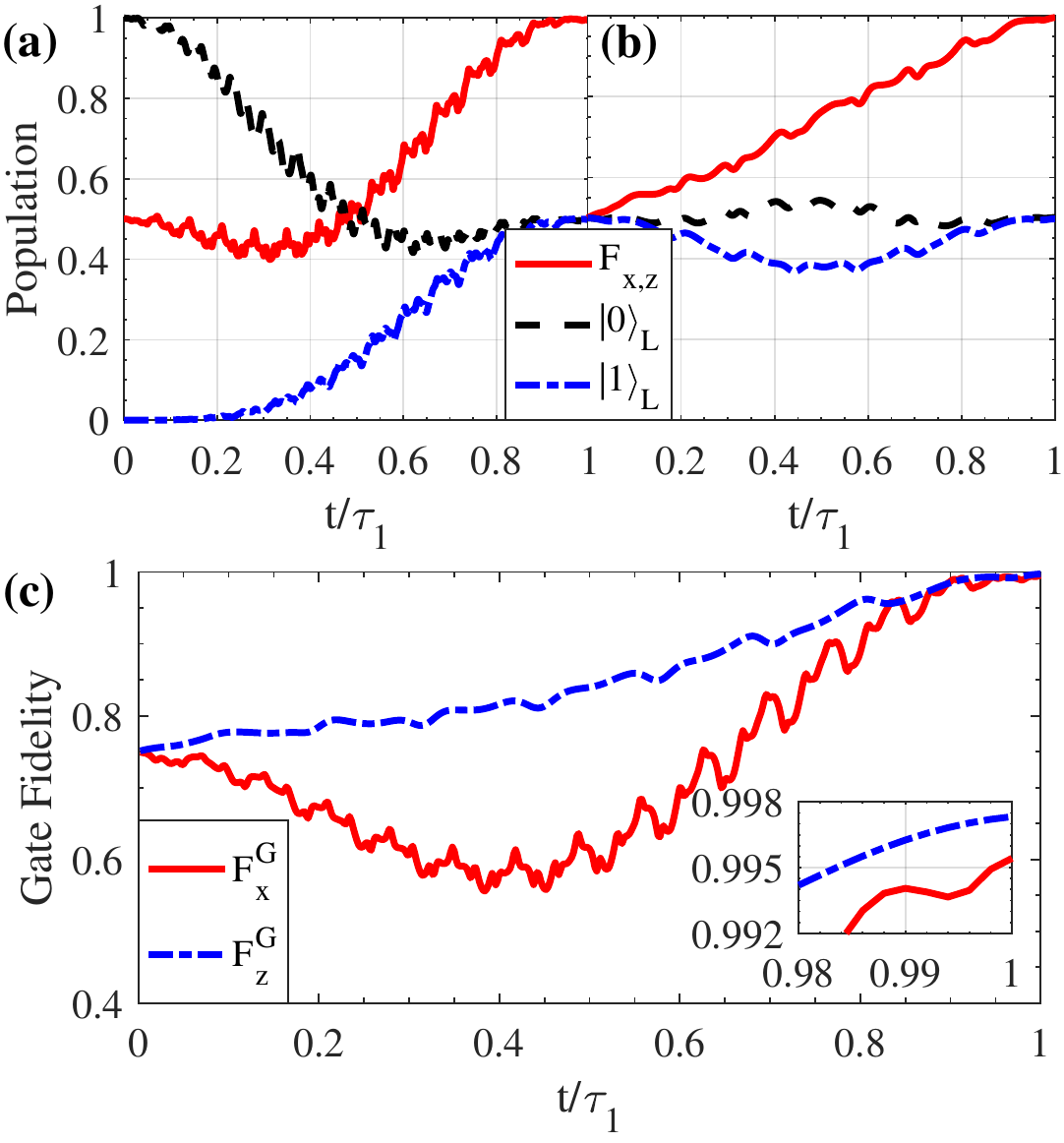}
  \caption{The logical-qubit-state population and fidelity dynamics of (a) $R^\textrm{T}_x(\pi/2)$ and (b) $R^\textrm{T}_z(\pi/2)$ with initial prepared states being $|0\rangle_L$ and $(|0\rangle_L+|1\rangle_L)/\sqrt{2}$, respectively. (c) Dynamics of the gate fidelities of $R^\textrm{T}_x(\pi/2)$ (dotted line) and $R^\textrm{T}_z(\pi/2)$ (solid line).}\label{Fig3}
\end{figure}

To check the validation of our scheme under realistic conditions, we next consider the effects of decoherence and high-order oscillating terms to further analyze our gate performance, the quantum dynamics of which can be simulated by the Lindblad master equation \cite{ME}. In our simulation, on the basis of the state-of-art experiments \cite{Martinis14}, we set decay and dephasing rates of different transmons to be identical as $2\pi\times4$ kHz. The intrinsic anharmonicities of transmon are $\alpha_1=2\pi\times 320$ MHz, $\alpha_2=2\pi\times 300$ MHz and the coupling strength is $g_{_{12}}=2\pi\times 12$ MHz, respectively. Take $R^\textrm{T}_x(\pi/2)$ and $R^\textrm{T}_z(\pi/2)$ as two typical examples, which can be obtained by modulating $\beta_2\approx1.43$ and $0$ to ensure $\theta=\pi/2$ and $\pi$. Supposing the single-logical qubit is initially at states $|0\rangle_L$ and $(|0\rangle_L+|1\rangle_L)/\sqrt{2}$, we evaluate these two gates using the state fidelity defined by $\textrm{F}_{x,z}=\langle\psi_{f_{x,z}}|\rho_1|\psi_{f_{x,z}}\rangle$ with the target ideal final states being $|\psi_{f_x}\rangle=(|0\rangle_L-\textrm{i}|1\rangle_L)/\sqrt{2}$ and $|\psi_{f_z}\rangle=(|0\rangle_L+e^{\textrm{i}\frac{\pi}{2}}|1\rangle_L)/\sqrt{2}$, respectively, where $\rho_1$ is the solved density matrix of the considered single-logical-qubit system by the Lindblad master equation. State population and fidelity dynamics of $R^\textrm{T}_x(\pi/2)$ and $R^\textrm{T}_z(\pi/2)$ are shown in Figs. \ref{Fig3}(a) and \ref{Fig3}(b), resulting in the final state-fidelities can reach $\textrm{F}_x=99.64\%$ and $\textrm{F}_z=99.63\%$, under parameters $\beta_{1}\approx1.58$ and $1.98$, $\Delta=0$ and $2\pi\times 18$ MHz, respectively, with a same $\Delta_1=2\pi\times 500$ MHz. In addition, for a general initial state $|\psi_1\rangle=\cos\theta_1|0\rangle_L+\sin\theta_1|1\rangle_L$ with $|\psi_{f'_{x,z}}\rangle=R^\textrm{T}_{x,z}({\pi}/{2})|\psi_1\rangle$ being the ideal final states, we can also define gate fidelity as $\textrm{F}_{x,z}^\textrm{G}=\frac {1} {2\pi}\int_0^{2\pi} \langle \psi_{f'_{x,z}}|\rho_1|\psi_{f'_{x,z}}\rangle d\theta_1$ \cite{gatefidelity} with the integration is numerically done for 1001 input states with $\theta_1$ being uniformly distributed over $[0, 2\pi]$. The obtained gate fidelities of $R^\textrm{T}_x(\pi/2)$ and $R^\textrm{T}_z(\pi/2)$ can reach $99.51\%$ and $99.74\%$, respectively, as shown in Fig. \ref{Fig3}(c). Through our numerical analysis, we find that gate infidelities of $R^\textrm{T}_x(\pi/2)$ and $R^\textrm{T}_z(\pi/2)$ mainly come from the decoherence of qubit system and high-order oscillating terms, which are about $(0.29\%, 0.20\%)$ and $(0.12\%, 0.14\%)$, respectively. Moreover, our numerical simulation is based on the full Hamiltonian $\mathcal{H}^{\varepsilon}_{12}$ and does not rely on any approximation, and thus verifies our scheme.


We next turn to the two-logical-qubit holonomic control-phase gate with the TOC technique, which can be combined with implemented arbitrary single-logical-qubit holonomic gates to achieve universal time-optimal NHQC. In this case, as shown in Fig. \ref{Fig1}(a), we continue to select the two adjacent transmons $\textrm{T}_3$ and $\textrm{T}_4$ (or $\textrm{T}_{3'}$ and $\textrm{T}_{4'}$) to encode the second logical qubit. Thus, there exists a two-logical-qubit DFS, i.e., $S_2=\{|0202\rangle_{1234}=|00\rangle_L, |0220\rangle_{1234}=|01\rangle_L, |2002\rangle_{1234}=|10\rangle_L, |2020\rangle_{1234}=|11\rangle_L\}$ which can be controlled only by the tunable coupling between the two capacitive coupled transmons $\textrm{T}_2$ and $\textrm{T}_3$, where the transmon $\textrm{T}_3$ is modulated by a two-tone qubit-frequency driving, in the form of $\omega_3(t)=\omega_3+\varepsilon_3(t)+\varepsilon_4(t)$ with $\varepsilon_i(t)=\dot{\mathcal{F}}_i(t)$, where $\mathcal{F}_i(t)=\beta_i(t)\sin(\omega_{\varepsilon_i}t+\Delta_{\varepsilon_i}t +\phi_{\varepsilon_i}(t))$. Therefore, we next start from analyzing the coupling interaction of physical-qubit subspace $\{|00\rangle_{23},|02\rangle_{23},|20\rangle_{23},|22\rangle_{23}\}$, and then determine how two-logical-qubit DFS is manipulated. In the interaction picture, the transformed Hamiltonian reads as
\begin{eqnarray}
\label{EqH23E}
\mathcal{H}_{23}^{\varepsilon}(t)&&=
           \left\{ |02\rangle_{23}\langle 11|e^{\textrm{i}(\Delta_2-\alpha_3)t}+|11\rangle_{23}\langle 20|e^{\textrm{i}(\Delta_2+\alpha_2) t} \right. \notag\\
& &\left.+\sqrt{3}|13\rangle_{23}\langle 22|e^{\textrm{i}(\Delta_2+\alpha_2-2\alpha_3) t}\right. \notag\\
& &\left.+\sqrt{3}|22\rangle_{23}\langle 31|e^{\textrm{i}(\Delta_2+2\alpha_2-\alpha_3) t}\right\} \notag\\
& &\times  \sqrt{2}g_{_{23}} \prod_{i=3,4}\left[e^{\textrm{i}\beta_i(t)\sin(\omega_{\varepsilon_i}t +\Delta_{\varepsilon_i}t+\phi_{\varepsilon_i}(t))}\right]
+\mathrm{H.c.},\ \ \ \ \ \
\end{eqnarray}
where $\Delta_2=\omega_3-\omega_2$ is the qubit-frequency difference between transmons $\textrm{T}_3$ and $\textrm{T}_2$. By using the Jacobi-Anger identity expansion, and then modulating qubit-driving frequencies to meet $\Delta_2+\alpha_2-(\omega_{\varepsilon_3}+\omega_{\varepsilon_4}) =-(\Delta_2-\alpha_3-\omega_{\varepsilon_3})=\Delta'$ with $\Delta'\ll\{\omega_{\varepsilon_3}, \omega_{\varepsilon_4}\}$, see Ref. \cite{sm} for details, Eq. (\ref{EqH23E}) under the rotating-wave approximation can form a tunable three-level structure with the effective coupling strength $g_3(t)$ and $g_4(t)$ in the subspace $\{|02\rangle_{23}, |20\rangle_{23}\}$ with an auxiliary subspace $\{|a\rangle_{L_2}=|11\rangle_{23}\}$. However, different from the single-logical-qubit case, we here have to consider the coupling of the state $|22\rangle_{23}$ to the non-computational subspace $\{|13\rangle_{23},|31\rangle_{23}\}$, the corresponding energy level diagram as shown in Fig. \ref{Fig1}(c), which can also form a tunable three-level structure with the effective coupling strength $\sqrt{3}g_3(t)$ and $\sqrt{3}g_4(t)$.

Specifically, for the dynamic process within the subspace $\{|02\rangle_{23}, |11\rangle_{23},|20\rangle_{23}\}$, under the unitary transformation  \cite{sm}, it  also forms a detuned coupling between the dressed-state $|\psi_+\rangle_{L_2}=\cos\frac{\vartheta}{2}|02\rangle_{23} +\sin\frac{\vartheta}{2}e^{-\textrm{i}\varphi}|20\rangle_{23}$ and $|a\rangle_{L_2}$ with effective strength $g'(t)=\sqrt{g_3^2(t)+g_4^2(t)}$ as
\begin{eqnarray}
\label{EqH23eff}
\mathcal{H}_{L_2}(t)=-\frac{\Delta'}{2}\tilde{\sigma}^z_{L_2} +g'(t)\left(e^{-\textrm{i}\varphi_3(t)}|\psi_+\rangle_{L_2}\langle a|+\mathrm{H.c.}\right),\ \ \ \
\end{eqnarray}
where the operator $\tilde{\sigma}^z_{L_2}=|\psi_+\rangle_{L_2}\langle \psi_+|-|a\rangle_{L_2}\langle a|$, parameters $\vartheta=2\tan^{-1}[g_4(t)/g_3(t)]$ and $\varphi=\varphi_4(t)-\varphi_3(t)$; while state $|\psi_-\rangle_{L_2}=\sin\frac{\vartheta}{2}|02\rangle_{23} -\cos\frac{\vartheta}{2}e^{-\textrm{i}\varphi}|20\rangle_{23}$ is decoupled. Meanwhile, similar to the single-logical-qubit case, we then ensure that $g'(t)$ and $\dot{\varphi}_3(t)$ both to be constant, that is $g'(t)=g'$ and $\dot{\varphi}_3(t)=\eta_2$, to meet the restriction of TOC. Therefore, by setting $\xi'(\tau_2)=\int_0^{\tau_2}\sqrt{g'^2+(\Delta'+\eta_2)^2/4}dt=\pi$ and tuning $\beta_4=0$ to make $\vartheta=0$ at the final time $\tau_2$, the  state $|02\rangle_{23}$ will be accumulated an unconventional holonomic phase as $e^{\textrm{i}\gamma_2}|02\rangle_{23}\langle 02|$ with $\gamma_2=\pi- \eta_2\tau_2 /2$.

\begin{figure}[tbp]
  \centering
  \includegraphics[width=\linewidth]{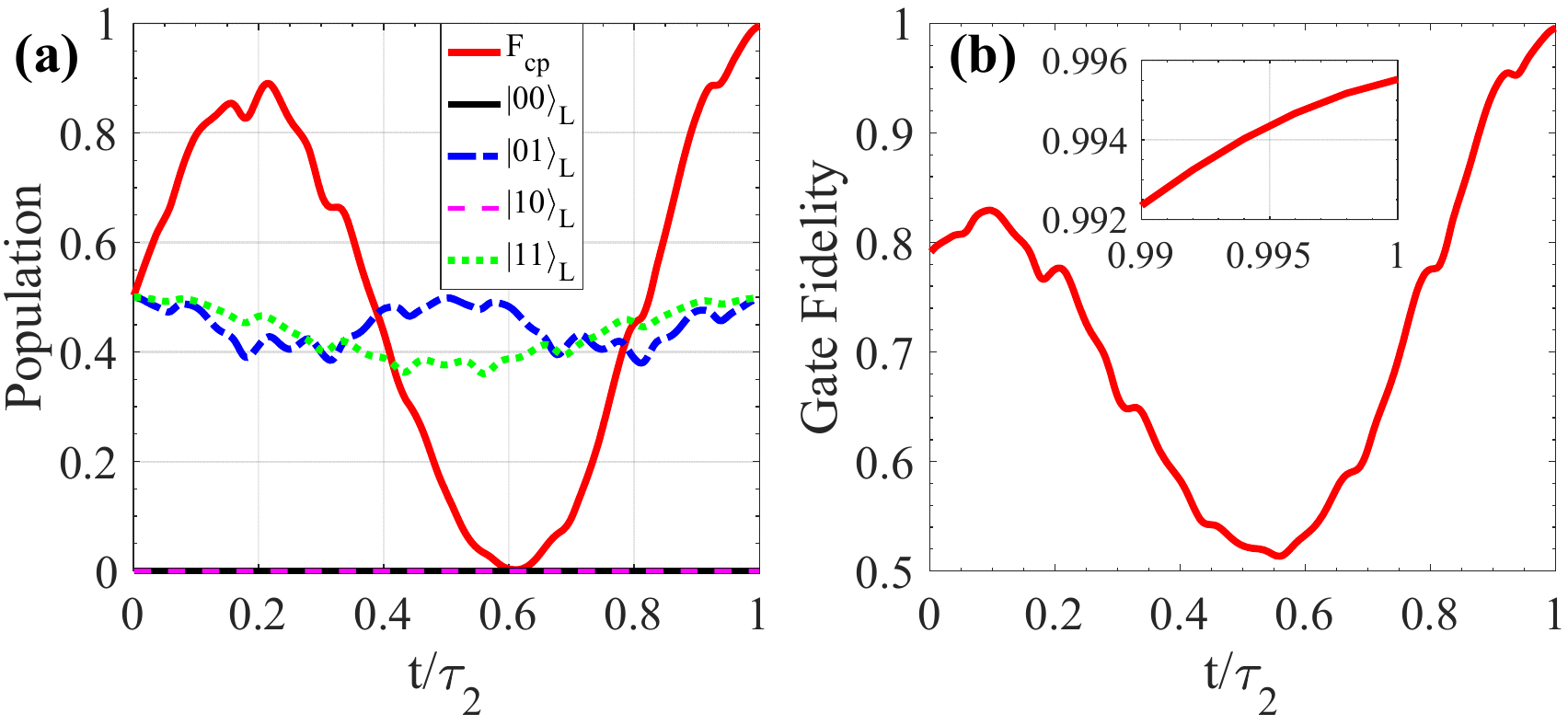}
  \caption{Performance of two-logical-qubit holonomic control-phase gate. The logical-qubit-state population and fidelity dynamics of $U_{L_2}(\pi/2)$ with initial prepared state being $(|01\rangle_L+|11\rangle_L)/\sqrt{2}$. (b) Dynamics of the gate fidelity of $U_{L_2}(\pi/2)$.}\label{Fig4}
\end{figure}

Meanwhile, for the leakage coupling of the state $|22\rangle_{23}$ into the non-computational subspace $\{|13\rangle_{23},|31\rangle_{23}\}$, under the premise of $\beta_4=0$, we move to the rotating frame defined by $U_\textrm{R}=\exp[-\textrm{i}\frac{\Delta'+\alpha_3-\alpha_2+\eta_2}{2}(|22\rangle_{23}\langle 22|-|13\rangle_{23}\langle 13|)t]$ and then assure $\xi_3(\tau_2)=\int_0^{\tau_2}\sqrt{3g'^2+(\frac{\Delta'+\alpha_3-\alpha_2+\eta_2}{2})^2}dt=2\pi$ to
satisfy the cyclic evolution condition, the physical-qubit state $|22\rangle_{23}$ will be cyclically evolved back to itself, leaving the operation to be identity, and thus effectively suppress unwanted leakage out of $S_2$. In this way, the time-evolution operator within $S_2$ is
\begin{eqnarray}
\label{EqUL2}
U_{L_2}(\gamma_2)=\text{diag}(1, 1, 1, e^{\textrm{i}\gamma_2} )
\end{eqnarray}
at the optimal time $\tau_2=\tau'_0\sqrt{1-(1+\Delta'/\eta_2)^2(1-\gamma_2/\pi)^2}$, which is also faster than conventional holonomic operations \cite{SingleloopSQ,Singleloop} with the gate time being $\tau'_0=\pi/g'$.

We next take an example of $\gamma_2=\pi/2$ to evaluate our gate performance numerically. Here, we also set decay and dephasing rates of different transmons to be identical as $2\pi\times4$ kHz. The anharmonicity of the transmon and coupling strength are set to be $\alpha_3=2\pi\times 330$ MHz, $\alpha_4=2\pi\times 310$ MHz and $g_{_{23}}=2\pi\times 10$ MHz, respectively. Supposing two-logical qubit is initially prepared in state $(|01\rangle_L+|11\rangle_L)/\sqrt{2}$,
the corresponding state population and fidelity dynamics are shown in Fig. \ref{Fig4}(a), where the state fidelity can reach $\textrm{F}_{\textrm{cp}}=99.50\%$ under parameters $\Delta_2=2\pi\times 560$ MHz, $\beta_3=1.54$, and  $\Delta'\approx2\pi\times 11.8$ MHz. To fully evaluate this gate performance, for a general initial state $|\psi_2\rangle=(\cos\vartheta_1|0\rangle_L+\sin\vartheta_1|1\rangle_L)\otimes
(\cos\vartheta_2|0\rangle_L+\sin\vartheta_2|1\rangle_L)$ with $|\psi_{f'_{\textrm{cp}}}\rangle=U_{L_2}(\pi/2)|\psi_2\rangle$ being ideal final state, we define $\textrm{F}^\textrm{G}_{\textrm{cp}}=\frac {1} {4\pi^2}\int_0^{2\pi} \int_0^{2\pi} \langle \psi_{f'_{\textrm{cp}}}|\rho_2|\psi_{f'_{\textrm{cp}}}\rangle d\vartheta_1d\vartheta_2$ as the gate fidelity, where the numerical integration is done for 10001 input states with $\vartheta_1$ and $\vartheta_2$ uniformly distributed over $[0, 2\pi]$. As shown in Fig. \ref{Fig4}(b), the $U_{L_2}(\pi/2)$ gate fidelity  can reach $99.55\%$, where the decoherence from all four transmons and  high-order oscillating terms induced  gate infidelity are about $0.33\%$ and $0.12\%$, respectively.


In summary, we have proposed a scheme to implement time-optimal NHQC in the DFS on a 2D transmon lattice, with the minimal qubit resource and only two-body interaction, improving the main disadvantage of the previous NHQC schemes. Therefore, it provides a promising method towards high-fidelity and robust quantum computation.

\bigskip

\acknowledgments

We thank B.-J. Liu and Prof. Luyan Sun for helpful discussions. This work was supported by the Key-Area Research and Development Program of GuangDong Province (Grant No. 2018B030326001), the National Natural Science Foundation of China (Grant No. 11874156), the National Key R\&D Program of China (Grant No. 2016YFA0301803), and the Innovation Project of Graduate School of South China Normal University (Grant No. 2019LKXM006).

\end{document}